# An image cytometer based on angular spatial frequency processing and its validation for rapid detection and quantification of waterborne microorganisms

Juan Miguel Pérez[a], Marc Jofre[a], Pedro Martínez[a], Adela Yañez Amoros[b], Vicente Catalan Cuenca[b], and Valerio Pruneri[a, c]

We introduce a new image cytometer design for detection of very small particulate and demonstrate its capability in water analysis. The device is a compact microscope composed of off-the-shelf components, such as a light emitting diode (LED) source, a complementary metal–oxide–semiconductor (CMOS) image sensor, and a specific combination of optical lenses that allow, through an appropriate software, Fourier transform processing of the sample volume. Waterborne microorganisms, such as *Escherichia coli* (*E. coli*), *Legionella pneumophila (L. pneumophila)* and *Phytoplankton*, are detected by interrogating the volume sample either in a fluorescent or label-free mode, i.e. with or without fluorescein isothiocyanate (FITC) molecules attached to the micro-organisms, respectively. We achieve a sensitivity of 50 cells/ml, which can be further increased to 0.2 cells/ml by pre-concentrating an initial sample volume of 500 ml with an ad-hoc fluidic system. We also prove the capability of the proposed image cytometer of differentiating microbiological populations by size with a resolution of 3 µm and of operating in real contaminated water.

## Introduction

Early detection is critical for the effective treatment and prevention of diseases. In many cases, diagnosis requires time-consuming, costly procedures and instruments, thus limiting their use to centralized settings with relatively advanced infrastructures and well-trained healthcare professionals[1]. Waterborne diseases cost the health care system over $500 million annually in the US and a similar amount in Europe; investments in early detection could thus lead to prevention of diseases and significant healthcare cost savings[2]. In both developing and industrialized nations, a growing number of contaminants from human activities are entering water supplies [3]; parasitic infections and diarrheal diseases caused by waterborne bacteria have become a leading cause of malnutrition owing to poor digestion of the food eaten by people sickened by water[4]. In recent years, for example, the pathogenic role of *Escherichia coli (E. coli)* has increased, as it can cause a variety of infections in humans[5]. *E. coli* is the most frequent cause of bloodstream infections and urinary tract infections, causes neonatal meningitis and is responsible for many food-borne infections worldwide[6]. In this study we detect and characterize waterborne microorganisms, such as *Escherichia coli*, *Legionella pneumophila* and *phytoplankton*, with the aim of reaching early identification and diagnosis of serious diseases. We also detect and characterize *Saccharomyces cerevisiae* microorganisms. *E. coli* and *S. cerevisiae* are the most widely studied prokaryotic and eukaryotic model organisms[7]; furthermore *S. cerevisiae* is a type of yeast used extensively in food production processes[8], therefore its detection and control have significant industrial interest. *L. pneumophila* is one of the main causes of severe atypical pneumonias in humans and it is present in soil, natural and man-made aquatic environments[9]. The *phytoplankton* microorganisms detected were saltwater diatoms which play an important role in ballast water contamination[10].

Within the food industry, bacterial culture methods are considered the gold standard by the food and drug administration (FDA)[5]. Lens-based epifluorescence microscopes, standard equipment in biological imaging, are also used for detection of micro-organisms[11]. However, the need of cost-effective tools with a shorter response time led to the development of several technologies, such as flow cytometry, polymerase chain reaction (PCR) and enzyme linked immunosorbent assay (ELISA)[12-16]. Besides these, intensive research has focused on the development of new biological sensors for rapid detection and identification of microorganisms, which can be classified into four main groups, depending on the transduction mechanism: optical, mass, electro-chemical and thermal[17]. Among those, optical biosensors have the highest share (35%)[18].

Due to their steady advances, complementary metal-oxide-semiconductor (CMOS) and charged coupled device (CCD) image sensors have great potential in lowering prices of optical biosensors by substituting sophisticated microscopes with simpler proximity detection schemes. This also allows replacing expensive laser sources with more economic light emitting diodes (LEDs). Image cytometers (I-CYTs) use CMOS or CCD to capture cells located on a microscope slide or in a transparent chamber and analyze thousands of them at once[19] For example, a chip-scale fluorescent I-CYT has been demonstrated reaching a spatial resolution of 10μm over a 24 mm$^2$ field of view (FOV)[15]. A more complex lensless fluorescent I-CYT with a FOV of 60mm$^2$ and spatial resolution below 4 μm has been designed using a fiber optic faceplate[21-22]. An I-CYT accessory has been also applied to a cell-phone for fluorescence imaging[23-26]. Besides resolution and FOV, the depth of field (DOF) is another important parameter to quantify the capability of a microscopy system, including that of an I-CYT. Microlens arrays (MLAs) have been used to increase DOF by multi-view point imaging[27] and for fluorescence quantification[28].

In this work, we introduce a new design of I-CYT composed of off-the-shelf components and capable of analyzing particulate within a sample volume of 350 μl, in a single snap-shot. In any imaging system, the sample is illuminated by an incident light beam. After the sample a lens can be placed which forms the image plane and the Fourier plane. In classical imaging the detector is placed at the image plane, where a replica of the sample (an image) can be captured. Instead, in the proposed I-CYT the detector is placed at the Fourier plane, so that the detected image is the Fourier transform in the spatial frequency domain (FT) of the sample (Fig.1). Furthermore, the transmitted beam is spatially filtered in a set of sub-images through MLA just before being detected by a CMOS image sensor, this resulting in increased FOV and DOF. Note that imaging systems that analyze the data in the Fourier domain have already been reported[29-31]. However, the new design of the proposed I-CYT makes it capable of retrieving the sample information without the need of lengthy and complex digital transformations with high computational cost. By combining the I-CYT with an appropriate fluidic system to concentrate the sample, we demonstrate label-free detection of specific microorganisms in water with a concentration as low as 0.2 cells/ml. The platform can also identify different populations by analyzing them in size and complexity. As we will see in the next sections, fluorescence labelling is an option for increasing further the differentiation capability (specificity).

## Designs and Methods

### Design of the image cytometer (I-CYT)

The sample volume is illuminated by an incoherent light source. The interaction between the sample and the incident beam results in the emission of beamlets with a broad range of spatial frequencies. After traversing an optical lens, the complex light field distribution is proportional to the FT of the beam which has interacted with the sample[31-32]. The angular spatial frequency information contained in the FT can be inverted, revealing the structure of the sample in the space domain[33].

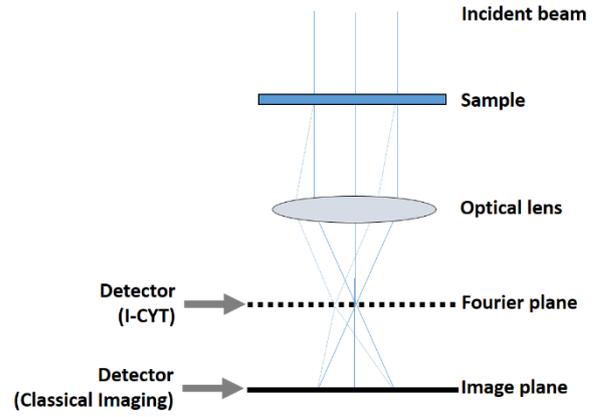

Fig. 1: comparison between classical imaging and proposed I-CYT; the location of the detector for both schemes is indicated.

Data extraction is based on Fourier optics principles. The center of the detected pattern is the zero spatial frequency component while the n-th order harmonic component is located at a distance $n\lambda Z_i f_0$ from the center, where λ is the wavelength, $Z_i$ the distance from the lens to the MLA and $f_0$ the lens's focal length. From the statistical analysis of the captured sub-images, one can determine essential parameters, such as particulate distribution (size and counting) and complexity; the latter being proportional to the absorption of the particles within the sample. The size and complexity distributions can be used to achieve specificity and differentiate particle populations, as it will be clear in the following sections of the paper. To further increase the specificity, fluorescein isothiocyanate (FITC) labeling can be used. This requires selecting a specific pump wavelength with a proper filter and at the same time suppressing the residual pump with an additional filter before the fluorescence signal is detected. FITC has an excitation/emission spectrum centered at 495 and 519 nm, respectively. For the fluorescence detection the polarizers are not required.

Retrieving sample information through data in the spatial frequency domain instead of the space domain results in a system capable of analyzing large volumes (350 μl) in a single capture, thanks to the combination of a large FOV (>24 mm$^2$) and DOF (≈1 mm). The large FOV is achieved by placing the sample in close proximity to the optical lens which results in a magnification factor of approximately one. The increased DOF is achieved by using the multi-aperture MLA to spatially filter the frequency information and propagate it towards the CMOS.

The information processing capacity of the proposed optical system and its sensitivity can be analyzed in terms of the spatial bandwidth product (SBP). Fourier holograms and their influence on the SBP have been extensively studied; the required SBP for a detector to ensure no loss of sample information is given by equations 1 and 2 below[34]. As shown, the SBP of a Fourier hologram depends on the wavelngth (λ), the distance of the sample from the detector ($d_1$) and the maximum recorded spatial frequency or its inverse value which is the minimum resolvable feature $\delta_{min}$.

$$SBP = \lambda \cdot d_1 \cdot 2 \cdot v_{max}^2 \qquad (1)$$

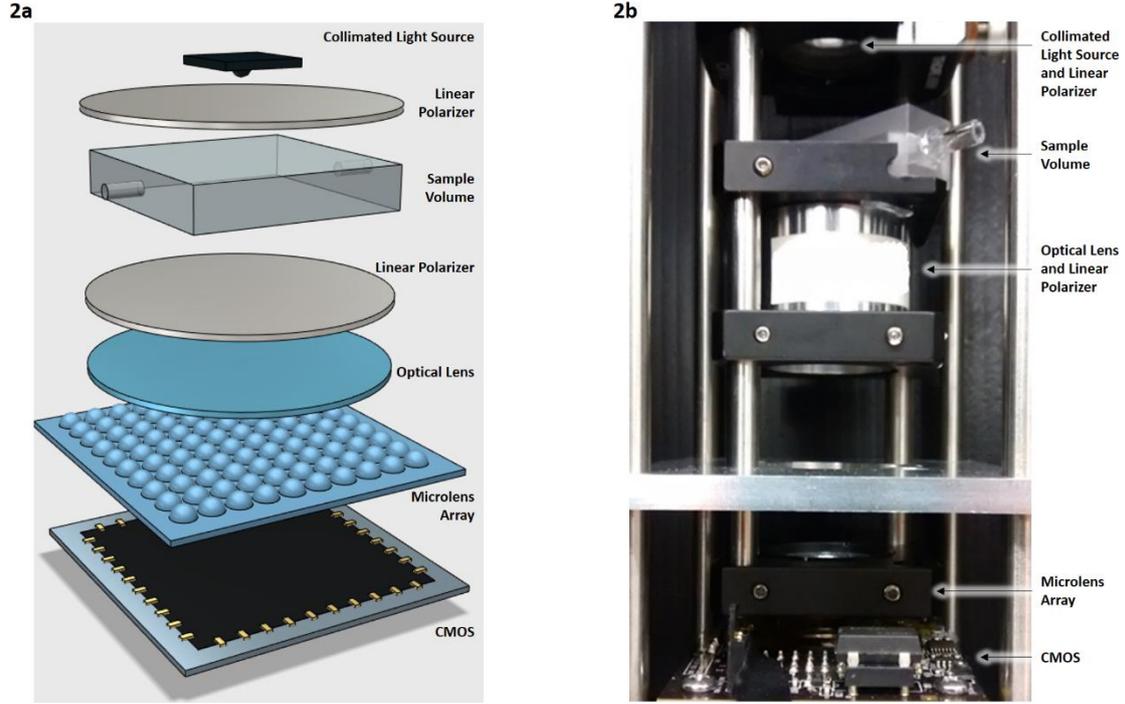

Fig. 2: Schematic of the I-CYT (Fig. 2a) and photograph of the laboratory prototype (Fig. 2b). It is composed of a light source, a flow cell chamber that contains the sample volume, an optical transforming lens, a microlens array and a CMOS image sensor. Two polarizers are placed, respectively, before and after the sample volume to generate a cross-polarized dark field image. The distance from the light source to the CMOS is about 95 mm.

$$SBP = \lambda \cdot d_1 \cdot \left(\frac{1}{\delta_{\min}}\right)^2 \quad (2)$$

Equation 1 is valid under the assumption of an ideal lens as a transforming element. For our optical system, the finite lens aperture results in the attenuation of high frequerncy components known as vignetting effect[35]. Equations 3 to 5 describe this vignetting effect in which the attenuation of the spatal frequencies (v) depends on the size of the illuminated sample (A), the dimension of the lens ($D_L$) and the distance between sample and lens ($d_2$). Equation 3 describes the conditon for absence of specturm attenuation, Equation 4 partial attenuation and 5 complete attenuation.

$$|v| < \frac{D_L - A}{2 \cdot \lambda \cdot d_2} \quad (3)$$

$$\frac{D_L - A}{2 \cdot \lambda \cdot d_2} < |v| < \frac{D_L + A}{2 \cdot \lambda \cdot d_2} \quad (4)$$

$$|v| > \frac{D_L + A}{2 \cdot \lambda \cdot d_2} \quad (5)$$

Fig. 2a shows the schematic of the system while Fig. 2b the laboratory prototype. It includes: a red, green and blue (RGB) fiber coupled LED light source emitting at wavelengths 465 nm, 515 nm and 635 nm, a quartz flow cell, an optical lens with focal length of 30 mm, an MLA with an area of 10x10 mm$^2$ and a pitch of 300 μm, two ultrahigh contrast linear polarizers with a maximum cross polarized transmission of 0.02%, 0.018% and 0.015% at 450, 550 and 650nm, respectively; finally, a 5 Mega pixel (MP) CMOS color image sensor with 2 μm pixel size is used for detection; the Bayer filter (RGB array of pixels) of the CMOS image sensor enables color detection. The MLA is placed from the lens at a distance equal to its focal length to discretize the FT in sub-images, detected in specific regions of the CMOS placed in proximity to the MLA. To enhance the detection signal, two cross polarizers can be placed on each side of the sample. In this way the light that comes from the source and does not interact with the sample will not reach the CMOS. When FITC labelling is used, the light source is substituted by a free space LED light with emission at 470 nm followed by a pump filter centered at 466 nm, with 40nm bandwidth and 60dB extinction ratio; a fluorescent filter, centered at 520 nm, with 36 nm bandwidth and 60 dB extinction ratio, is placed between the lens and the MLA.

With the described system specifications and the mathematical modeling of Equations 1 to 3, one can infer that the proposed I-CYT has the capacity to detect and acquire information of particles as small as 0.98 μm.

The CMOS image sensor of the laboratory prototype had a dimension of 5.70x4.28mm$^2$, which combined with the MLA allowed for the detection of 221 sub-images containing spatial frequency information of the captured sample. From each sub-image, the particulate size, complexity and count are computed, as it is explained in the following. Positioning the MLA over the CMOS is achieved by means of a Z-translating mechanical piece aligned and attached to the CMOS sensing area, thus minimizing XY positioning errors. The adjustment on the Z axis is done as to create non-overlapping sub-images (avoid aliasing) without resorting in sub-sampling (information loss). The optimum Z location of the MLA is at the microlens focal length.

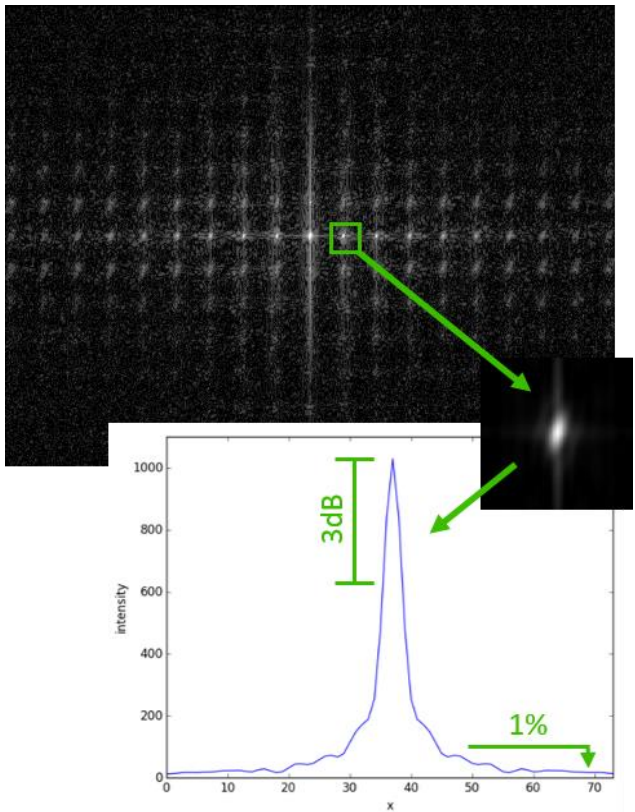

Fig. 3: graphical representation of the pattern analysis using a real capture from the I-CYT. The recovered pattern is composed by a set of sub-images and each sub-image is analyzed separately to obtain size information from the main lobe at its 3dB point, the complexity by contrasting the maximum (Imax) to the lowest 1% luminance and the count by integrating the pixel intensities.

The LED source is collimated using an achromatic lens and a graduated iris diaphragm. The collimation of the light source is of special importance as it enables uniform illumination. However, in practice, the sample is illuminated with a Gaussian wavefront; this implies lower signal to noise ratio for high spatial frequencies, with a direct effect on resolution. To partially counteract this effect, after capturing the sample volume image, the first processing steps are normalization and noise removal, which are accomplished by using a reference capture and a dark noise capture, respectively.

The reference capture is the image with no sample in place, while the dark noise capture with the light source turned off. Next, the interrogated sample is de-convoluted with a control sample for compensation. The control is a sample with zero particulate and preferably in the same buffer solution as that of the interrogated sample. Once the de-convolution is performed, the recovered sample is divided into the 221 sub-images and processed to retrieve the complexity, size and count information.

The complexity parameter is a measure of particulate absorption. It is calculated by analyzing the contrast of the sub-image. The sub-image maximum intensity (Imax) is contrasted against the mean value of the lowest 1% luminance of the complete image (the composition of all 221 sub-images). The size information is calculated by analyzing the patterns in each sub-image. A main lobe region is detected and its area is measured. The main lobe corresponds to the largest pattern within the sub-image with the area measured at the 3 dB point (where the intensity of the lobe is at half maximum). Finally the counting of the particulate is achieved by integrating and averaging the pixel intensities of each sub-image. For the label-free detection, the presence of particulate within the sample volume results in scattering of the incoming beam which is detected by the image sensor; for the fluorescent detection, the target microorganism is excited and emits at the corresponding fluorescence wavelength captured by the CMOS image sensor. These pixel intensities from either scattering or fluorescence emission have a linear response, after the analog to digital conversion of CMOS image sensor's response. The linear measurements can be transformed into a logarithmic scale to provide results in cells/ml by calibrating the system with a serial dilution. Fig. 3 shows a graphical representation of the pattern analysis.

The use of spatial and temporal incoherent light reduces collective (coherent) effects from particle clusters in the detected images, thus permitting a more efficient particle distinction. At the same time, short sample acquisition time minimizes further the effect on measurements that would be associated to the formation of said clusters.

**Sample preparation**

To prepare spiked samples, *L. pneumophila* and *E. coli* were grown at 37ºC on Buffered Charcoal Yeast extract medium (BCYE) and on Tergitol 7 agar for 3 days and 24h hours, respectively. Cells were harvested, a Phosphate Buffered Saline (PBS) and 10-fold serially diluted suspensions in PBS were prepared.

A second serial 10-fold dilution of E.coli was prepared, but in this case using water from a cooling tower as diluent, and samples were treated with an anti-*E. coli* O + *E. coli* K (FITC) antibody by abcam[36] for fluorescent detection.

All dilutions were prepared inside a Class II biosafety cabinet using sterile materials, and volumes of 2 ml were loaded into the quartz flow cells to be measured with the I-CYT. Prepared samples were also cultured following the conditions and methods previously described in order to know the microorganism's concentration in the samples.

Sample volumes of *S. cerevisiae* and *phytoplankton* were also prepared and captured with the I-CYT. *S. cerevisiae* are yeast microorganisms of about 5 µm in size which were used to test the label-free differentiation. The *phytoplankton* microorganisms were salt water diatoms with a known size of approximately 14 µm; they have a natural fluorescence emission (autofluorescence) at a wavelength of 610 nm when excited at 488 nm; since the excitation wavelength spectra of FITC and autofluorescence overlap, one is able to capture at the same time and differentiate FITC-labeled *E. coli* and autofluorescent *phytoplankton*. The diatoms were provided by Marine Eco Analytics (MEA-NL)[37].

**Fluidic system for sample concentration**

A fluidic system was designed to enhance the performance of the I-CYT and detect low levels of concentration. The fluidic system comprises a barometric pump and a hollow fiber membrane filter CellTrap$^{TM}$ to concentrate the microorganisms suspended in a large volume and elute them into the 2ml volume. Three low concentration samples were prepared using the fluidic system as explained next.

First, 1 ml volumes of $10^4$, $10^3$ and $10^2$ cells/ml original *E. coli* in PBS dilutions were added into independent screw cap bottles with deionized water for a total volume of 500 ml; resulting in *E. coli* concentrations of 20, 2 and 0.2 cells/ml. The bottles were sealed with a pressurized cap with an input from the barometric pump and an output towards the CellTrap$^{TM}$ membrane filter. The filter comprises a set of 0.2 µm hollow fiber membranes which trap the particulate passing through and free the filtered water down one end; once the complete volume has passed the filter, all the particulate will be trapped inside the membranes and can be eluted into a chosen volume using a luer tip syringe. After concentration and elution to 2 ml volumes, the concentrations of the three samples become 5000, 500 and 50 cells/ml. Fig. 4 displays the fluidic system on both of its stages (concentration and elution). Fig. 4a shows the concentration step: the initial volume leaving a bottle, passing through the CellTrap$^{TM}$ hollow fibers before the waste is collected in a second bottle. Fig. 4b shows the elution process in which a syringe is used to recover the trapped microorganisms into a 2 ml volume.

**Results and Discussion**

We demonstrated the capabilities of the proposed I-CYT by measuring and analyzing a variety of waterborne microorganisms (*E. coli, L. pneumophila, phytoplankton* and *S. cerevisiae*) over a wide range of concentrations, in both label-free and fluorescent modes. The samples were prepared according to the protocols previously described.

The detection of the microorganisms can be conveniently represented using a dispersion graph - size versus complexity (Figs. 5a, 5b, 6a and 6b) - and a concentration graph – size versus concentration (Fig. 6c). Each point on the dispersion graphs corresponds to a single sub-image. The processing of the signal as described above allows to resolve concentrations of different waterborne microorganisms and differentiate them from a control sample of PBS or a white buffer (microbiological charge zero). The accuracy of the concentration measurements using the proposed I-CYT are within 0.5 log as it is estimated by comparing them with nominal data and microbiological culture results (Figs. 5c, 5d, 5e, 5f). Figs. 5d and 5e also display the correlation coefficients (r) between the I-CYT concentration and the microbiological culture with correlations of 0.9966 and 0.9386 calculated by Pearson's correlation coefficient.

Fig. 5a displays the label-free detection of a *L. pneumophila* serial dilution in PBS and a water sample taken from cooling towers, contrasted to a control sample in buffer solution. The water sample had an unknown amount of biological and inert material and the label-free design permits a complete analysis of its microbiological charge. The water from the cooling towers was then spiked with FITC-labeled *E. coli* microorganisms. Fig. 5b shows the corresponding fluorescent detection results where FITC-labeled *E. coli* microorganisms are characterized in complexity and size, similarly to the *L. pneumophila* samples. Thanks to the fluorescent labelling and related optical filtering, the *E. coli* microorganisms can be also differentiated from debris. Fig. 5c displays the counting of *L. pneumophila* of Fig. 5a, with each dilution measured twice; two series, 1 and 2, together with their average (series Avg). In all cases the measured value on the vertical axis can be compared with the nominal concentration of the sample on the horizontal axis.

Fig. 5d shows the concentration of FITC-labeled *E. coli* compared to measurements obtained by standard microbiological culture of the captured dilutions. Fig. 5 e displays results after the sensitivity enhancement through the fluidic CellTrapTM concentrator. The initial volume of 500 ml was reduced to 2ml by maintaining the same amount of biological charge. This improved the concentration detection limit of the entire system from about 50 to 0.2 cells/ml. Fig. 5f is the zoom of a specific region of Fig. 5e, from which one can easily differentiate 0.2, 2 and 20 cells/ml. These results prove the capability of the proposed I-CYT to quantify microbiological charge contained in 500 ml volume with a single measurement.

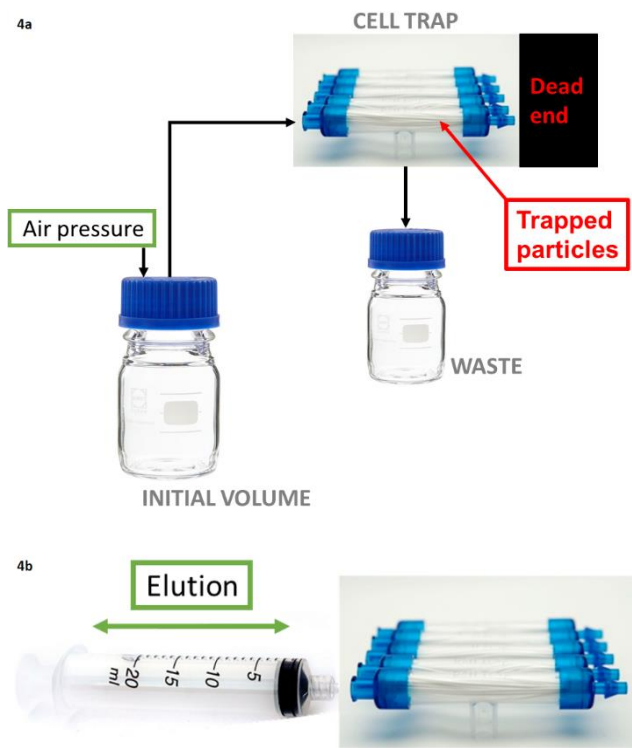

**Fig. 4: Schematic of concentrator fluidic system. Fig 4a shows the concentration step and the bottle with the initial volume passing through the CellTrap$^{TM}$. Fig 4b shows the elution process in which a syringe is used to recover the trapped microorganisms in 2 ml volume.**

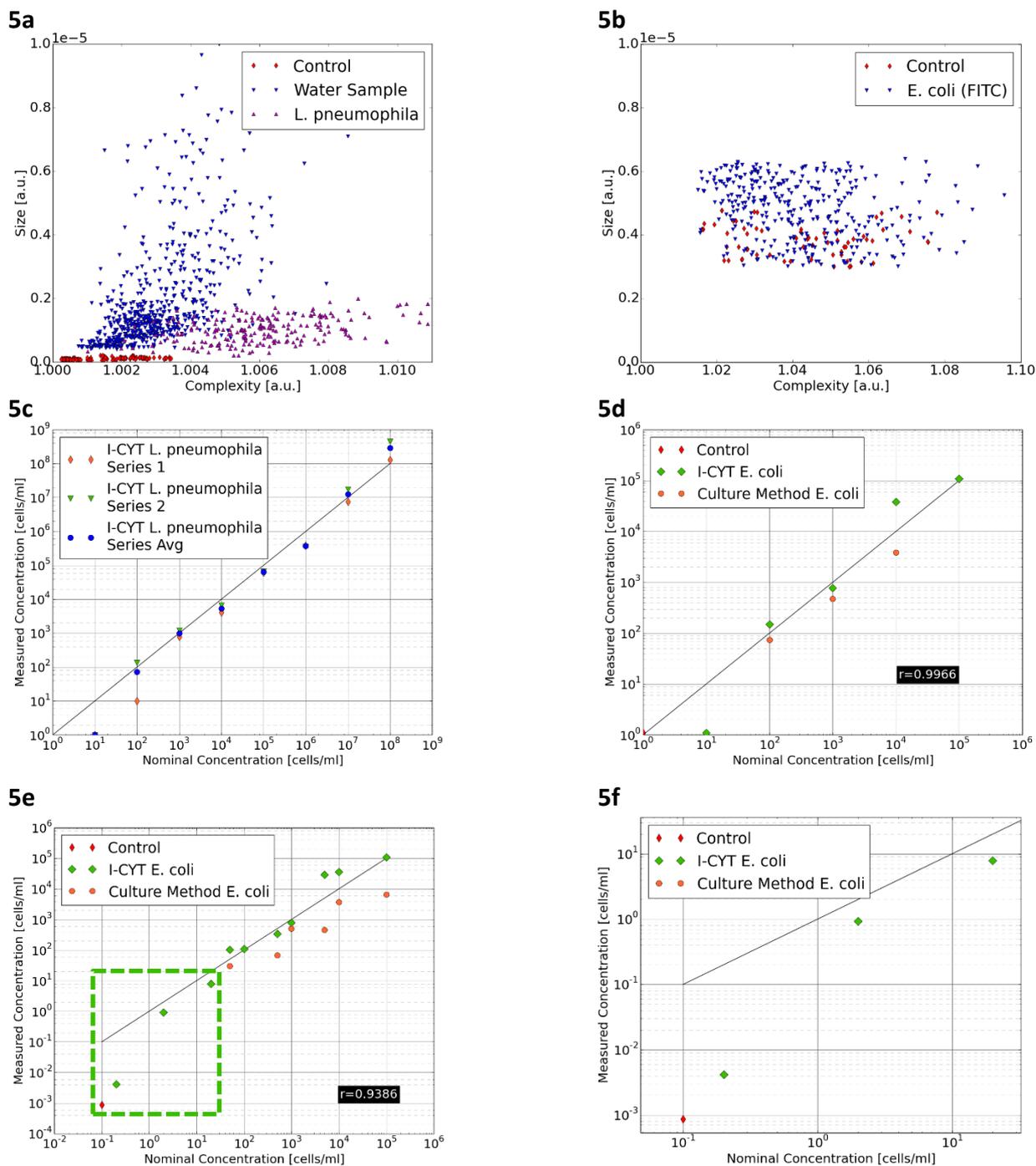

Fig. 5: detection and quantification of waterborne microorganisms with the proposed I-CYT. Fig. 5a displays the label-free detection of an *L. pneumophila* serial dilution in PBS and a water sample taken from cooling towers; Fig. 5b the fluorescent detection of the FITC-labeled *E. coli* suspended in the contaminated cooling tower water sample; Fig. 5c the counting of *L. pneumophila* of Fig. 5a; Fig. 5d the counting of the FITC- labeled *E. coli* of Fig. 5b; Fig. 5e results after sensitivity enhancement through the fluidic CellTrapTM concentrator; Fig. 5f the zoom of a specific area. All the concentration measurements (Figs. 5d, e and f) are compared to standard microbiological culture.

We also demonstrated the capability of the system to identify microbiological populations, with a difference in size larger than 3 μm (about the pixel size). This was achieved in both label-free

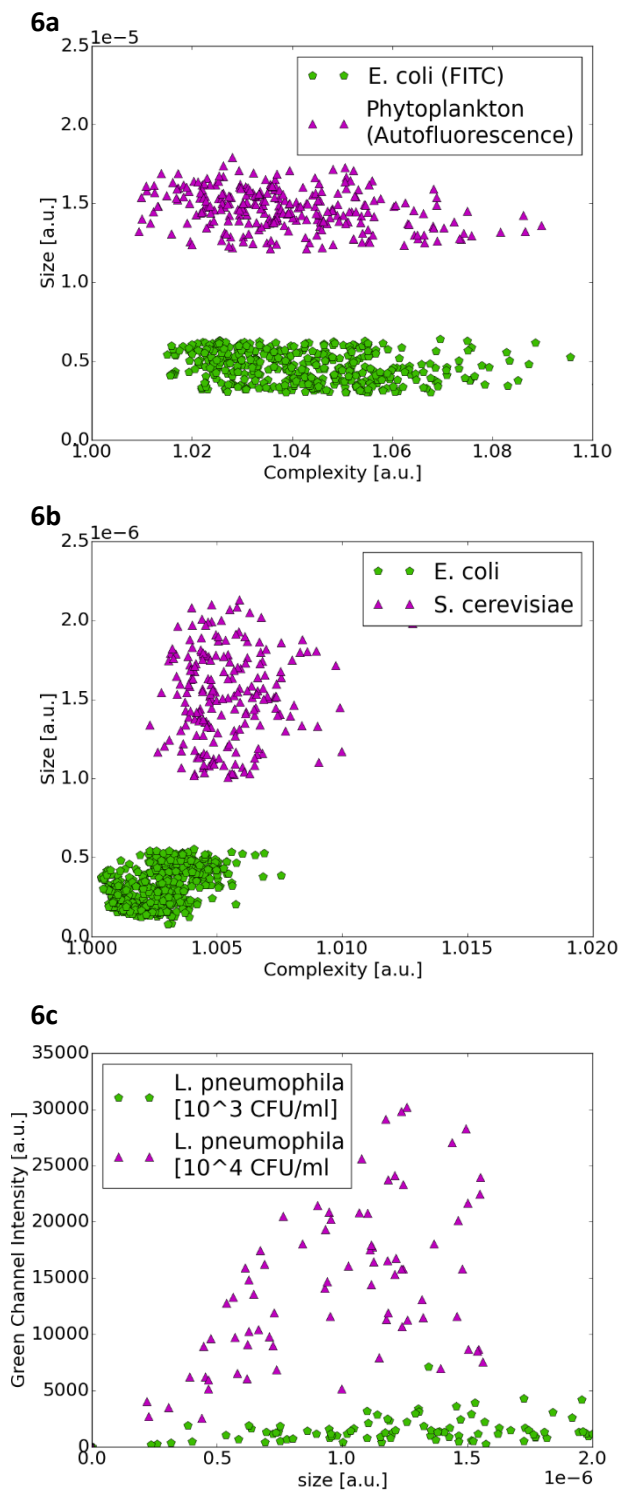

**Fig. 6: fluorescent and label-free population differentiation. Fig. 6a** shows complexity versus size graph of FITC-labeled *E. coli* compared to *phytoplankton* **Fig. 6b** the label-free differentiation between *E. coli* and *S. cerevisiae*. **Fig. 6c** size versus intensity comparing two *L. pneumophila* samples of different concentrations.

and fluorescent capturing of *E. coli, L. pneumophila, S. cerevisiae* and *phytoplankton* (Fig.6). *E. coli* dilutions were labeled with FITC emitting at 519nm, whereas *phytoplankton* autofluorescence at 610 nm. Differentiation of these two fluorescent emissions in our I-CYT simply consists in placing the filter for the targeted fluorescence signal and processing the image using the RGB components of the CMOS sensor with the Bayer filter. By processing the Red channel signal, the *phytoplankton* microorganisms can be detected and analyzed in complexity and size, while the Green channel can do the same for the FITC labeled microorganisms (*E. coli* in our case). Fig. 6a displays the complexity versus size of FITC-labeled *E. coli* microorganisms compared to *phytoplankton*. *E. coli* microorganisms are known to have a size around 2 µm, while the *phytoplankton* around 14 µm. It is clear that these two micro-organism populations can be easily differentiated by the proposed I-CYT,

Using the I-CYT in label-free configuration, unlabeled *E. coli* and *S. cerevisiae* were detected, processed and analyzed. *S. cerevisiae* is a type of yeast with an average particle size of 5 µm. Fig. 6b shows the complexity and size parameters for the two samples which clearly indicate how the I-CYT allow their differentiation despite their close dimensions. Samples with smaller difference in size (<3 µm) can only be differentiated by their concentration (captured intensity). For example Fig. 5c shows intensity versus size dispersion for two *L. pneumophila* samples with different concentration ($10^3$ cells/ml and $10^4$ cells/ml).

## Conclusions

We have introduced a new design of I-CYT which is based on the detection of the Fourier transform image of the sample plane. The sample capture and analysis in the spatial frequency domain results in an increased FOV (>24mm$^2$) and DOF (≈1mm). When combined with an ad-hoc fluidic system these allow to reach detection levels of 0.2 cells/ml and analyze hundreds of milliliters of sample in a single capture. In addition, the capacity of differentiating populations by size, complexity and fluorescence emission are highly suitable for multi-target analysis, while the label-free capability for rapid detection of microorganisms. We have demonstrated the potential of the proposed I-CYT by identifying, quantifying, differentiating waterborne microorganisms of high environmental and societal interest, such as *E. coli*, *phytoplankton*, *L. pneumophila*, and the most widely used yeast, *S. cerevisiae*. The newly designed I-CYT is portable, low-cost, built with off-the-shelf components and highly suitable for rapid and point-of-care detection of micro-organisms, cells and other particulates. It thus has the potential to replace high-end and large instrumentation with comparable sensitivity and large dynamic detection (concentration) range, especially where in situ and immediate countermeasures are required. This is the case, for example, of remote areas which lack infrastructures and laboratories where analysis are typically performed.

## Acknowledgements

This work was financially supported by Spanish Ministry of Economy and Competitiveness (MINECO), "Fondo Europeo de Desarrollo Regional" (FEDER) through grant TEC2013-46168-R, NATO's Public Diplomacy Division in the framework of "Science for Peace" and by Fundació Privada CELLEX.

## Notes and references

[a] ICFO-Institut de Ciencies Fotoniques, Mediterranean Technology Park, 08860 Castelldefels (Barcelona), Spain.

[b] LABAQUA S.A., c) Del Dracma, 16-18, 0314, Alicante, Spain.

[c] ICREA - Institució Catalana de Recerca i Estudis Avançats, Barcelona, Spain.